\documentclass[prd, amsfonts, twocolumn, nofootinbib, showpacs]{revtex4}
\usepackage{graphicx, epsfig}
\usepackage{color}
\usepackage{amsmath}
\usepackage[normalem]{ulem}
\newcommand{\be}{\begin{equation}}
\newcommand{\ee}{\end{equation}}
\newcommand{\bea}{\begin{eqnarray}}
\newcommand{\eea}{\end{eqnarray}}

\newcommand{\gapp}{\mathrel{\raise.3ex\hbox{$>$}\mkern-14mu
\lower0.6ex\hbox{$\sim$}}}
\newcommand{\lapp}{\mathrel{\raise.3ex\hbox{$<$}\mkern-14mu
\lower0.6ex\hbox{$\sim$}}}
\def\bbox{{\,\lower0.9pt\vbox{\hrule \hbox{\vrule height 0.2 cm
\hskip 0.2 cm \vrule  height 0.2 cm}\hrule}\,}}

\usepackage{color}

\begin{document}
\title{ Emergent Spacetime in Stochastically Evolving Dimensions}
\author{Niayesh Afshordi$^{1,2,3}$, Dejan Stojkovic$^{1,3}$}
\affiliation{$^1$ Perimeter Institute for Theoretical Physics, 31 Caroline St. N., Waterloo, ON, N2L 2Y5, Canada}
\affiliation{$^2$ Department of Physics and Astronomy, University of Waterloo, Waterloo, ON, N2L 3G1, Canada}
\affiliation{ $^3$ HEPCOS, Department of Physics, SUNY at Buffalo, Buffalo, NY 14260-1500, USA}


\begin{abstract}
Changing the dimensionality of the space-time at the smallest and largest distances has manifold theoretical advantages.
If the space is lower dimensional in the high energy regime, then there are no ultraviolet divergencies in field theories,
it is possible to quantize gravity, and the theory of matter plus gravity is free of divergencies  or renormalizable.
If the space is higher dimensional at cosmological scales, then some cosmological problems (including the cosmological constant problem)
can be attacked from a completely  new perspective. In this paper, we construct an explicit model of ``evolving dimensions" in which the dimensions
open up as the temperature of the universe drops. We adopt the string theory framework in which the dimensions are fields that live on the string worldsheet,
and add temperature dependent mass terms for them. At the Big Bang, all the dimensions are very heavy and are not excited. As the universe cools down, dimensions open up one by one. Thus, the dimensionality of the space we live in depends on the energy or temperature that we are probing.  In particular, we provide a kinematic Brandenberger-Vafa argument for how  a discrete {\it causal set}, and eventually  a continuum $(3+1)$-dim spacetime  along with Einstein gravity emerge in the Infrared from the worldsheet action. The $(3+1)$-dim Planck mass and the string scale become directly related, {\it without any} compactification. Amongst other predictions, we argue that LHC might be blind to new physics even if it comes at the TeV scale. In contrast, cosmic ray experiments, especially those that can register the very beginning of the shower, and collisions with high multiplicity and density of particles, might be sensitive to the dimensional cross-over.
\end{abstract}


\maketitle
\section{Introduction}

We believe that (with some exceptions) we understand our universe on scales approximately between $\sim 10^{-17}$ to $10^{27}$ cm's. The first scale corresponds to the energy scale of TeV which is the energy probed in the highest energy accelerators available so far. The second scale corresponds to the distance characteristic for super-clusters of galaxies, i.e. the scale at which cosmology kicks in. At scales shorter than $10^{-17}$ cm and larger than about a Gpc  $\sim 10^{27}$ cm, we are running into problems.

There exists a strong motivation to reduce the dimensionality of the space-time at high energies (short distances). One of the most acute problems - the Standard Model hierarchy problem does not exists in $(1+1)$-dim space-time since the corrections to the Higgs mass are only logarithmically divergent. There is no need for new particles and elaborate cancelation schemes. The coupling constant in QCD in $(1+1)$ and $(2+1)$ dimensions has positive dimension, and the theory is therefore super-renormalizable, i.e. only a finite set
of graphs need overall counter terms. Even, the most elusive concept in modern physics - quantum gravity - is much more within reach in lower dimensions. If the fundamental short scale physics is lower dimensional, there is no need to quantize $(3+1)$-dim gravity. Instead we should quantize  $(2+1)$ and $(1+1)$ dimensional gravity, which are, by comparison, much easier tasks.  $(2+1)$-dim general relativity has no local gravitational degrees of freedom, i.e. no gravitational waves in classical theory and no gravitons in quantum theory. Gravity is then completely determined by the local distribution of masses. The number of degrees of freedom in such a theory is finite, quantum field theory reduces to quantum mechanics and the problem of non-renormalizability disappears \cite{Carlip:1995zj}. For the reason of simplicity, $(1+1)$ dimensional gravity is even more attractive. Einstein's action in $(1+1)$  dimensional space-time is a topological constant (Euler's characteristic of the manifold in question) and the theory is trivial (unless augmented by some additional fields). Models of gravity in $(1+1)$ dimensions are completely solvable \cite{Klosch:1997md,LouisMartinez:1993cc} and considerable work has been done on their quantum aspects \cite{Grumiller:2006ja,Zaslavskii:2003eu,Giddings:1992ae,Callan:1992rs,Bogojevic:1998ma}.

On intermediate scales between $10^{-17}$ cm and a Gpc, we know pretty well that our space is three dimensional. However, there is some  motivation to change the dimensionality of the space-time on larger scales, comparable to the present cosmological horizon. Such ideas have been explored in a class of  brane-world models, known as {\it cascading gravity} \cite{Gregory:2000jc,Dvali:2000hr,deRham:2007xp,Hao:2014tsa,Dai:2014roa}. Even more explicit constructions that address the cosmological constant problem from a completely new perspective were introduced in \cite{Anchordoqui:2010er}. If the forth spatial dimension opens up at the current horizon scale, then an effective cosmological constant of the correct magnitude is induced without putting it into the equations by hand.

 On the experimental front, it is very intriguing that some evidence for the lower dimensional structure of our space-time on a TeV scale might already exist. Namely,
alignment of the main energy fluxes in a target (transverse) plane has been observed in families of cosmic ray particles in high altitude cosmic ray experiments \cite{et1986,Mukhamedshin:2005nr,Antoni:2005ce} (high altitude is crucial in order to catch the very beginning of the shower before the energies significantly degrade). The fraction of events with alignment is statistically significant for families with energies higher than TeV and large number of hadrons. This can be interpreted as evidence for co-planar scattering of secondary hadrons produced in the early stages of the atmospheric cascade development. Explicitly, co-planar scattering would then indicate that the fundamental physics above the TeV scale is $2+1$-dimensional rather than $3+1$-dimensional. On the other end of the energy spectrum, possible observational evidence for cosmological-scale extra dimensions were discussed in \cite{Afshordi:2008rd}.

Lower dimensional scattering has very important predictions for the Large Hadron Collider (LHC) physics.  There are three consequences which should be observable if the physics becomes planar at the TeV scale: {\it (i)\/} cross-section of hard scattering processes changes compared to that in the SM as the momentum transfer becomes comparable with the crossover scale; {\it (ii)\/} $2 \to 4$ and higher order scattering processes at high energies become planar, resulting, {\it e.g.\/}, in four-jet events, where all jets are produced in one plane in their center-of-mass frame, thus strikingly different from standard QCD multijet events; {\it (iii)\/} under certain conditions, jets of sufficiently high energy may become elliptic in shape (for details see \cite{Anchordoqui:2010er,Anchordoqui:2010hi,Stojkovic:2013lga}). It is also important to note that no new fundamental particles are expected to exist in order to solve the hierarchy problem.

Another distinct prediction of such a dimensional reduction scheme comes from the nature of gravity in lower dimensions.  It is well-known that, in a $(2+1)$-dim general relativity, there are no local gravitational degrees of freedom, and hence there are no gravitational waves (or gravitons).  If the universe was indeed $(2+1)$-dim at some earlier epoch, it is reasonable to deduce that no primordial gravitational waves of this era exist today.  There is thus a maximum frequency of primordial gravitational waves, implicitly related to the dimensional transition scale, beyond which no waves can exist.  This indicates that gravitational wave astronomy can be used as a tool for probing this scale \cite{Mureika:2011bv}.

From the model building point of view,  a framework of ``evolving" or ``vanishing" dimensions was proposed in \cite{Anchordoqui:2010er} in which the space at scales shorter than $10^{-17}$ cm is lower dimensional, while at scales larger than a Gpc it is higher dimensional. In this setup, the number of dimensions increases with the length scale.  On the shortest distances at which our space appears as continuum, the space is one-dimensional. At a certain critical length scale, the space becomes effectively two-dimensional. At the scale of about $10^{-17}$ cm, the space becomes effectively three-dimensional. Finally, at the scales of about a Gpc, the space becomes effectively four-dimensional. In a dynamical picture where the universe starts from zero size and then grows, the dimensions open up as the universe expands and temperature drops. The {\it ad hoc} model that was used in this proposal was an ordered lattice, which captures all the basic features of the proposal and allows one to make generic model independent predictions. However, so far no explicit model in terms of fundamental Lagrangians was constructed.

The main goal of this paper is to construct an explicit model of ``evolving dimensions". To do so, we will use the existing apparatus of the string theory and modify it to
achieve the change of dimensionality of our space with the energy scale.

\section{Stringy Model}\label{sec:stringy}

We start from the standard Nambu-Goto action:
\be \label{nga}
S_{\rm free} = -\frac{1}{2 \pi \alpha '} \int d^2 \xi \sqrt{-\gamma}
\ee
where $\alpha'^{-1}$ is the string tension, and $\gamma$ is the determinant of the metric on the string world sheet  $\gamma_{ab}$
\be \label{gamma}
\gamma_{ab} = g_{\mu \nu} \partial_a X^\mu \partial_b X^\nu .
\ee
The metric in the target space $g_{\mu \nu}$ is usually considered to be fundamental and  $\gamma_{ab}$ induced. However, we will adopt the {\it opposite} view here. The lower dimensional metric $\gamma_{ab}$ will be considered fundamental, and higher dimensional manifold $g_{\mu \nu}$ induced since it is woven by an evolving lower dimensional submanifold (as in Fig.~\ref{distance}).
Coordinates on the string world-sheet are $\xi^a = (\tau, \sigma)$. The coordinates in the target space are $X^{\mu}(\tau, \sigma)$. The index $\mu = (0,1,2,3, ... , n)$ where $n$ is the number of space-like dimensions in the target space. $X^{\mu}$ represent the coordinates in the space that we live in (we do not fix the dimensionality of that space {\it a priori}).
We can also understand $X^{\mu}$ as fields that live on the string world-sheet, so in principle, we can add mass terms for them. This would break the conformal symmetry in the IR, which would require more careful standard string theory interpretation. However, for the purpose of a phenomenological theory with desired properties, conformal symmetry is not crucial.

We will now illustrate the basic idea by adding temperature dependent masses for the fields $X^{\mu}$
\bea \label{lmass}
&& L_{\rm mass} = m_0^2 e^{-m_1/T} X^1 X_1 +
 m_0^2e^{-m_2/T} X^2 X_2 + \nonumber \\ && m_0^2e^{-m_3/T} X^3 X_3 + \ldots + m_0^2e^{-m_n/T} X^n X_n  \nonumber
\eea
where $m_1 \gg  m_2 \gg m_3 \gg \ldots \gg m_n$.
The time-like coordinate $X^0$ is massless, so it is always excited (this assumption can easily be changed). Note that we can always use the gauge freedom to identify the time coordinate on the worldsheet with the time coordinate in the target space, which will result in identifying the temperature on the worldsheet with that in the target space (apart from a possible redshift factor if the string is moving).  Parameter $m_0$ has units of mass, and is of the order of the fundamental energy scale (perhaps $M_{Pl}$). We start from a hot Big Bang when $T \gg m_1, m_2, m_3, \ldots , m_n$, so all the fields are massive and are not excited. When the temperature drops to $T \ll m_1$, only the first field $X^1$ is practically massless and gets excited; when the temperature drops to $T \ll m_2$, only  $X^1$ and $X^2$ are excited, and so on. Today, at
$T \sim 10^{-3}{\rm eV} \ll m_3$, the first three fields are excited. In principle, the process does not have to stop at any finite number, so decrease in energy would be opening more and more dimensions.

The action corresponding to the mass Lagrangian (\ref{lmass}) is
\be \label{smass}
S_{\rm mass} = \frac{1}{2 \pi \alpha '} \int d^2 \xi \sqrt{-\gamma} L_{\rm mass}.
\ee
Note that fields $X^\mu$ have the physical interpretation as dimensions and have units of length, which makes the action (\ref{smass}) dimensionless.

The equations of motion that govern the excitations of the fields $X^\mu$ are
\bea  \label{eom1}
&& \left( \Box +   m_0^2 e^{-m_1/T} \right) X^1  = 0 \\
&& \left( \Box +   m_0^2 e^{-m_2/T} \right) X^2 = 0 \nonumber  \\
&& \ldots  \nonumber
\eea
where $\Box =\frac{1}{\sqrt{-\gamma}} \partial_a \left(\sqrt{-\gamma} \gamma^{ab} \partial_b   \right)$. Here we treated $X^{\mu}$ as Cartesian coordinates, while it is easy to add Christoffel symbols to the equations of motion  in case $g_{\mu \nu}$ in the target space is an arbitrary curved manifold.

Since all the components of the field multiplet $X^\mu$ have different masses, the general Lorentz (and diffeomorphism) invariance in the target space is broken. However, at any given temperature the Lorentz invariance is restored in some subset of the original $n$ dimensions, e.g. today at $T\sim 10^{-3}$eV the Lorentz invariance is effectively restored in the first three dimensions since all the first three fields are practically massless. This also implies that the usual string theory conformal invariance is restored in the subset of dimensions which are practically massless. This is important if one would like to preserve the straightforward string theory interpretation (which must include massless gravitons and gauge fields in its spectrum).

\subsection{Plasmon Dimensional Confinement}

The temperature dependent mass terms, similar to those in Eq. (\ref{lmass}) can generically be introduced via non-perturbative IR effects. For example, photons in a thermal electron-positron plasma at $T \lesssim m_e$ have a mass given by the plasma frequency:
\be
m_{\gamma}^2 =\omega^2_p = \frac{8\pi n_e \alpha_e}{m_e} = \left( 128 T^3 m_e\over \pi \right)^{1/2} \alpha_e e^{-m_e/T},\label{plasmon}
\ee
where $\alpha_e$ is the fine structure constant and $m_e$ is the electron mass. In this case, the collective interaction of photons with the plasma induces an effective mass, which becomes negligible at low temperatures, as the density of thermally produced electron-positron plasma vanishes exponentially for $T \ll m_e$.

 While Eq. (\ref{plasmon}) is for QED in $(3+1)$-dimensions, a similar phenomenon can happen for massless $X^\mu$ fields on the $(1+1)$-dim world sheet, if they couple to a thermal plasma of massive of excitations.

\subsection{Symmetron Dimensional Confinement}

A different possibility is to employ the class of {\it chameleon} or {\it symmetron} models \cite{Khoury:2003aq,Hinterbichler:2010es}. Namely, we can write down a microscopic theory where the fields actually couple to the energy density of the environment. For simplicity, let's consider only one of the fields, say $X^1$. A simple mass term in the form of $(1/2) \alpha' \rho m_0^2 X^1X_1$, where $\rho$ is the energy density of the environment, would suffice. Then, as the universe expands and energy density goes down, the mass of the field would decrease, as desired in our evolving dimensions framework. The challenge is to write down a microscopic theory where such coupling appears. This can be achieved by introducing the matter action in the form
\be\label{sym_action}
S_{m}=\int d^2 \xi  \sqrt{-\tilde{\gamma}} {\cal L}_m \left( \phi (\xi), \tilde{\gamma}_{ab}\right) ,
\ee
${\cal L}_m$ is the Lagrangian for the matter field, $\phi (\xi)$, which couples to the metric $\tilde{\gamma}_{ab}$ related to the original $\gamma_{ab}$ by the conformal factor $A(X^1)$
\begin{equation} \label{ej}
\tilde{\gamma}_{ab}=A^2(X^1)\gamma_{ab}
\end{equation}
The metrics $\gamma_{ab}$ and $\tilde{\gamma}_{ab}$ describe the Einstein and Jordan frames respectively.
From Eq.~(\ref{sym_action}), we have
\be
\frac{\delta S_m}{\delta X^1} = \frac{\delta S_m}{\delta \tilde{\gamma}^{ab}} \frac{\delta \tilde{\gamma}^{ab}}{\delta X^1}
\ee
Varying the kinetic term (\ref{nga}) for $X^1$ we get
\be
\frac{\delta S_{\rm free }}{\delta X^1} =   \frac{1}{2 \pi \alpha '}\partial_a \left(\sqrt{-\gamma} \gamma^{ab} \partial_b X^1  \right) = \frac{\sqrt{-\gamma} \Box X^1} 2 \pi \alpha '
\ee
Thus the equation of motion for the field $X^1$ in Jordan frame is
\be
 \frac{1}{2 \pi \alpha '}\Box X^1 =   \frac{1}{2} A^2,_{X^1} \tilde{T}
\ee
where $\tilde{T}$ is the trace of the matter stress-energy tensor in the Jordan frame, with
\be
\tilde{T}_{ab}=\frac{2}{\sqrt{-\tilde{\gamma}}}\frac{\partial \left(\sqrt{-\tilde{\gamma}}{\cal L}_m \right)}{\partial \tilde{\gamma}^{ab}},
\ee
 but the box operator is calculated with the metrics $\gamma_{ab}$.

The trace of the matter stress-energy tensor depends on the equation of state as $\tilde{T} = \tilde{T}^a_a = (1- w)\tilde{\rho}$ in $(1+1)$-dim, but for convenience we will define $\rho =A^{ 1+w} \tilde{\rho}$. This $\rho$ is now independent of $X^1$ for constant $w$, it is conserved in Einstein frame and has the usual redshifting properties (see e.g. \cite{Hinterbichler:2011ca}).
Thus, in Einstein frame, e.g. for a pressureless source $w=0$, the equation of motion is
\be
\frac{1}{2 \pi \alpha '}\Box X^1 = \rho  A,_{X^1} ,
\ee
which implies the effective potential
\be
V_{\rm eff} =  \rho A(X^1).
\ee
Then the concrete form of $V_{\rm eff}$ that gives the desired mass term can be obtained with a choice
\be
 A(X^1)=1+\frac{1}{2} m_0^2 X^1X_1 .
\ee
This choice gives an effective mass  for the field $X^1$ in the form of $ m_{\rm eff} = \alpha' \rho m_0^2$, as we desired.
Note however that the trace of the stress-energy tensor is classically zero for radiation (i.e. $\tilde{T} =0$ for $w=1$ in (1+1)-dim). For our model to work in the hot Big Bang scenario we have to provide the coupling to radiation as well.
We can always rely on quantum effects which will introduce the trace anomalies and provide coupling to radiation. A simple alternative is to add an extra factor in front of the Jordan frame Lagrangian
\be
S_m = \int d^2 \xi \sqrt{-\tilde{\gamma}} B(X^1)  {\cal L}_m \left( \phi , \tilde{\gamma}\right)
\ee
An implicit $X^1$-dependence of the Lagrangian would give the standard coupling to $T$ as before, but the new ingredient now is the $B(X^1)$ prefactor. This factor corresponds to a ``pressure" coupling. Its contribution to the equation of motion for $X^1$ is proportional to ${\cal L}_m$, which is the pressure. For example, for non-relativistic dust with zero pressure, it will not give any additional contribution. One of the concrete realizations consistent with all of the assumed symmetries is $B = 1+ m_0^2 X^1X_1$.

Thus the final equations of motion that govern the evolution of the fields $X^\mu$ are
\bea  \label{eom2}
&& \left( \Box + 2 \pi \alpha '  \rho m_0^2 \frac{m_1}{m_0} \right) X^1 =0  \\
&& \left( \Box + 2 \pi \alpha ' \rho  m_0^2 \frac{m_2}{m_0} \right) X^2 =0 \nonumber  \\
&& \ldots  \nonumber
\eea
where the scales $m_1 < m_2 < \ldots <m_n$ are inserted in the last step by hand in order to introduce the hierarchy in the number of dimensions which are excited at a given energy density of the universe. Strictly speaking, $\rho$ is the energy density of the mater field(s) $\phi(\xi)$ which live on the string worldsheet, however we mentioned that by choosing the gauge in which the time coordinate on the worldsheet is the same as in the target space, we can identify the temperature on the worldsheet with that in the target space (we will discuss more details about the transition from $\phi (\xi^a)$ to $\phi (X^\mu (\xi^a))$ in the next section). Thus, as the energy density (or equivalently temperature) in the universe drops, dimensions are excited one by one. Since energy density depends polynomially on the temperature, the dimensional cross-over is not as fast as the exponential one in Eq.~(\ref{lmass}).

 \section{Emergence of Locality: A worked example of $(1+1)$ to $(3+1)$-dim transition}
\label{el}

In the previous section, we introduced  a matter field $\phi (\xi^a)$ that lives on the string worldsheet. The string worldsheet itself evolves and builds a target space where the number of dimensions depends on the temperature. In this section we will examine the transition from $\phi (\xi^a)$ to $\phi (X^\mu (\xi^a))$. In particular we will start from the action for the field $\phi (\xi^a)$ which is not manifestly local, and arrive to the local action for the field $\phi (X^\mu)$ in the low energy limit.
This will resemble quantization of an effective field theory below its cut-off (for example a proton, or nucleus). This step is necessary in order to see what happens to the matter fields when we put them on a non-standard background of an evolving string.

\begin{figure}
   \centering
\includegraphics[width=7cm]{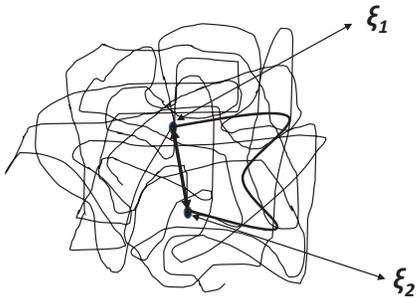}
\caption{An infinite string that intersects itself many times is fundamentally a $(1+1)$-dim object, but effectively builds a  $(3+1)$-dim structure on large distances.  The distance between two points on a string, $\xi^a_1$ and $\xi^a_2$, along the string is determined by the metric on the string $\gamma_{ab}$ defined in Eq.~(\ref{gamma}). However, the effective distance in an induced $(3+1)$-dim space between the same two points is given by the metric $g_{\mu \nu}$.}
\label{distance}
\end{figure}

An issue that needs careful attention in this step is the notion of distance in the target space spanned by $X^{\mu}$.
For simplicity, here we provide a concrete example of a $(1+1) \rightarrow (3+1)$-dim crossover. Short wavelength field excitations should see only a $(1+1)$-dim string, and thus travel along the string. However, long wavelength excitations effectively live on an induced  $(3+1)$-dim volume and do not see a short distance $(1+1)$-dim structure. The notion of distance in the  $(3+1)$-dim space must be then induced from the distance in the fundamental $(1+1)$-dim space (see Fig.~\ref{distance}).
In other words, we have to define the transition from $\phi (\xi^a)$ to $\phi (X^\mu (\xi^a))$ for some (scalar for simplicity) field that lives on such a structure.
This transition will be closely connected with two scales:
\begin{enumerate}
 \item The string coherence length scale, above which a $(1+1)$-dim string starts effectively looking like a $(3+1)$-dim volume.  We can characterize this scale through the typical extrinsic curvature of the string worldsheet ${\cal K}$.
 \item The length scale of non-locality, below which the multiple connectedness of the effective $(3+1)$-dim space becomes important. As we see below, this scale can be characterized through $N_c$, the density of intersections of a space-time lattice, or causal set, which approximates the continuum 3+1-dim space-time.
 \end{enumerate}

We will now use simple geometric arguments to  outline a dimensional cross-over mechanism.
%
%

 Consider a non-local contribution to the world-sheet action:
\be \label{action}
\Delta S_m= \int (\sqrt{-\gamma_1} d^2\xi_1)( \sqrt{-\gamma_2} d^2\xi_2) K\left(\Delta X^2 \right)  \left[\phi (\xi^a_1) - \phi (\xi^a_2)\right]^2,
\ee
where
\be
\Delta X^2 \equiv g_{\mu \nu} \left[ X^\mu(\xi^a_1) - X^\mu(\xi^a_2) \right] \left[ X^\nu(\xi^a_1) - X^\nu(\xi^a_2) \right],
\ee
and $K(\Delta X^2)$ is a scalar function of $\Delta X^2$.

 Now, in order to recover locality in the IR, we assume:
\bea
 K(\Delta X^2) =  \Box_4  \frac{\delta^4(\Delta X^\mu)}{\sqrt{-g}} + {\cal O}(\alpha') = \nonumber\\     \frac{1}{\sqrt{-g}} \frac{\partial }{\partial X^\mu}\left( \sqrt{-g} g^{\mu\nu} \frac{\partial}{\partial X^\nu} \frac{\delta^4(\Delta X^\mu)}{\sqrt{-g}}\right) + {\cal O}(\alpha'), \nonumber\\ \label{eq:Kdx}
\eea
 where $\alpha'$ corrections are expected due to quantum fluctuations of the string, which blur the points of intersection.

This term is necessary because the effective $(3+1)$-dim space is multiply connected if one goes along the string (there are many paths along the string between any two points in this space). To recover locality, only nearby points in target space should see each other, while other contributions should be highly suppressed for
$\Delta X^2 \gg \alpha'$. Now, using
\be
\phi(\xi_a) \simeq \frac{\partial \phi}{\partial X^{\mu}} \times \left[X^\mu(\xi_a) - X^\mu_0\right] + \phi_0,
\ee
for IR field fluctuations, we can do an integration by part to find:
\be  \label{action4d}
\Delta S_m \simeq \int (\sqrt{-g} d^4X) (N_c Z)  g^{\mu \nu} \, \frac{\partial \phi}{\partial X^{\mu}} \frac{\partial \phi}{\partial X^{\nu}},
\ee
where $N_c$ is the space-time density of string crossings, while $Z$ is a geometrical dimensionless factor defined as:
\be \label{z_def}
Z \equiv 8 \left\langle \int (\sqrt{-\gamma_1} d^2\xi_1)( \sqrt{-\gamma_2} d^2\xi_2) \frac{\delta^4(\Delta X^\mu)}{\sqrt{-g}}  \right\rangle.
\ee
averaged over individual string crossings. In the Appendix, we show that $Z$ is logarithmically divergent, and is approximately given by:
\be
Z \simeq  4\pi \log\left(2\over \epsilon\right),
\ee
where $\epsilon \sim {\cal K}\alpha'^{1/2}$ quantifies the extrinsic curvature of the strings over the string length scale $\alpha'^{1/2}$.

While the choice in Eq. (\ref{eq:Kdx}) for the non-locality kernel $K(\Delta X^2)$ appears {\it ad-hoc}, the above calculation demonstrates that terms more (less) divergent as $\Delta X \rightarrow 0$ lead to non-renormalizable (super-renormalizable) terms that vanish as power-laws in the UV (IR) cut-off. Therefore, on scales much bigger than the inverse density of the string network, but much smaller than the size of the system, we expect the Lorentz-invariant $(3+1)$-dim action (\ref{action4d}) to emerge.

\begin{figure}
   \centering
\includegraphics[width=8cm]{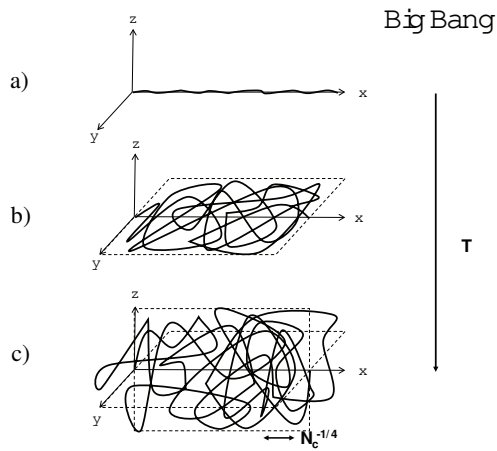}
\caption{Geometry of the string excitations. At temperatures much greater than the $(1+1) \rightarrow (2+1)$-dim crossover, i.e.  $T \gg m_2$ in Eq.~(\ref{lmass}), a string extends only in one dimension, say x, while all the other degrees of freedom are suppressed (time is not shown). At $m_3 \ll T \ll m_2$, a string extends in the xy plane by intersecting itself many times. At $m_4 \ll T \ll m_3$, a string covers the whole xyz space.  At $T \ll m_4$, a fourth spatial dimension should be opened up, but a string cannot effectively intersect itself in four spatial dimensions, so our universe remains $(3+1)$-dim at arbitrary low temperatures. The structure resembles a random  space-time lattice (or causal set) where the string intersections represent the lattice sites. The scale  $N_c^{-1/4}$  is determined by the geometry of the lattice and represents the length scale at which our space appears lower-dimensional today.  Note that the cross-over temperature in early universe does no have to coincide with $N_c^{1/4}$.}
\label{excitations}
\end{figure}

We now note that starting from a worldsheet $(1+1)$-dim action (\ref{action}), we have arrived at the continuum action for $(3+1)$-dim massless scalar field (\ref{action4d}) in the IR (up to a field renormalization). In fact, one can trace back this emergence of a continuum Lorentz-invariant IR action in 2 higher dimensions to a geometric or kinematic (as opposed to dynamical) realization of Brandenberger-Vafa mechanism \cite{Brandenberger:1988aj}, since string intersections are not generic in higher dimensions\footnote{Moreover, string intersections violate Lorentz symmetry in $(2+1)$ dimensions, as they persist in time, and thus can be thought of as a gas of particles that define a rest frame.} (see Fig.~\ref{excitations}).
In this concrete example, our universe will make an effective transition from $(1+1)$ to $(3+1)$ dimensions at a given temperature, but the geometry of the process would stop further opening of dimensions. In dimensions higher than $(3+1)$, strings would mostly miss each other, and a few rare intersections would not make a (quasi) continuous higher dimensional space. Thus, our universe would remain effectively $(3+1)$-dim at an arbitrary low temperature, despite the fact that more degrees of freedom are excited \footnote{ This conclusion can be altered if the fields $X^\mu (\xi^a)$ have a non-trivial potential with multiple minima. Then one can talk about the sequence of events. First, strings effectively build $(3+1)$-dim hypersurfaces (or branes), then these $(3+1)$-dim hypersurfaces build higher-dimensional hypersurafces by effectively intersecting themselves, and so on. Then the above explicit construction can be used to increase the effective IR dimension by increments of 2, successively.}.

 In the scenario that we described above,  we have an evolving string network, where intersections constantly appear and disappear as string segments cross each other.
One may also interpret these instantaneous string intersections as a {\it causal set} sprinkling of space-time (as opposed to only space) \cite{Bombelli:1987aa}, suggesting a potential connection between {\it causal set} and {\it string theory} approaches to quantum gravity. If this causal set picture is realized in nature, then not only the process of opening new dimensions would stop with $(3+1)$ dimensions, but the whole effective structure of space-time may disintegrate once the fourth spatial dimension is excited. The reason is again the fact that the string intersections are rare in $(4+1)$ dimensions, and those that were made earlier do not persist in time. In that case, instead of an effective higher dimensional description, a fundamental $(1+1)$-dim description will be forced back  at late times.

 Alternatively, it is possible that, due to the interaction term (\ref{action}), string network forms a 3-dimensional bound state (or condensate), that preserves its dimensionality even when a fourth spatial dimension opens up. In this case, the causal set lives on a $(3+1)$-dim membrane, that can freely move in higher dimensions. As an added bonus, one naturally expects an Einstein-Hilbert low energy effective action (and hence $(3+1)$-dim GR) for the membrane, simply on geometrical grounds. To see this, let us integrate out the field $\phi(\xi^a)$ for a fixed metric $g_{\mu\nu}(X)$:
\be
S_{geom.} \sim \int (\sqrt{-g} d^4X) (N_c Z) \langle \Box^*_4 \rangle \langle \phi^2 \rangle_{\rm UV},
\ee
where $\langle \Box^*_4 \rangle \sim \alpha'^{-1}$ depends on the effective width of the kernel $K(\Delta X^2)$, which regulates the derivatives at string intersections. $\langle \phi^2 \rangle_{\rm UV}$ is a UV divergent quantity which depends on the choice of vacuum state for $\phi$ field. However, under Lorentz-invariant regularizations of Hadamard states, only the logarithmic divergences remain, which using (\ref{action4d}) yields:
\be
N_c Z \langle \phi^2 \rangle_{\rm UV} \sim (\ln \Lambda) {\cal R},
\ee
where $\Lambda$ is proportional to the ratio of UV (i.e. $N^{1/4}_c$) to IR cut-off (or mass), while ${\cal R}$ is the $(3+1)$-dim Ricci scalar. Therefore, we find:
\be
S_{geom.} \sim \frac{\ln \Lambda}{\alpha'} \int (\sqrt{-g} d^4X) {\cal R},
\ee
which is Einstein-Hilbert action with the $(3+1)$-dim Planck mass:
\be
M^2_{p,4d} = (8\pi G_N)^{-1} \sim (\ln\Lambda)\alpha'^{-1}.
\ee
Surprisingly, the $(3+1)$-dim Planck mass and the string scale are now directly related, {\it without any} compactification in this framework! On the other hand, the cut-off of the $(3+1)$-dim effective description is given by $\Lambda_{\rm UV} \sim N_c^{1/4} \gtrsim $ TeV, as otherwise it should have already been probed by the LHC.

\section{Phenomenology}

In this section, we outline some possible phenomenological implications of the model that we have constructed here.

We start by re-emphasizing, that in our framework, dimensions as degrees of freedom open up consecutively at {\it world-sheet} temperatures determined by the mass parameters $m_i$ in the early universe. The resulting structure has its own geometric properties determined by space-time intersection density $N_c$. In general, $m_i$ and $N_c$ are unrelated, which means that the $(2+1) \rightarrow (3+1)$-dim crossover in early universe could have happened for example at $T\sim  m_3 = 10^{16}$GeV, but today a particle propagating on the  space-time lattice of density $N_c \sim$ (few TeV)$^{4}$ could in principle probe the lower dimensional structure already at the LHC.

Focussing again only on $(1+1) \rightarrow (3+1)$-dim crossover, for a finite tension string, the classical string network as in Figs.~\ref{distance}-\ref{excitations} has a finite dimensional crossover energy scale $N^{1/4}_c$. If quantum fluctuations only give negligible corrections to this picture, we practically have the ordered lattice model of vanishing dimensions introduced in \cite{Anchordoqui:2010er}. In such a model, the lattice is rigid and classical, and thus propagation of particles along the lattice links/blocks depends on the particle energy, i.e. short wavelength particles see $(1+1)$-dim string, while long wavelength particles see a $(2+1)$-dim surface and $(3+1)$-dim volume. For that case detailed predictions for the LHC physics were given in \cite{Anchordoqui:2010er,Anchordoqui:2010hi}.

The thermal and kinematic dimensional cross-over mechanisms developed in Secs. \ref{sec:stringy} and \ref{el} respectively, while complementary in our model, lead to distinct phenomenological implications.  The space-time lattice model implies that the number of dimensions that an individual particle sees changes with energy (wavelength) of that particle, what could in principle be tested in high energy particle collisions (like planar events, elliptic jets etc. \cite{Anchordoqui:2010er,Anchordoqui:2010hi}). In contrast, Eq.~(\ref{lmass}) and  Eq.~(\ref{sym_action}) imply that the number of dimensions changes with temperature, which requires finite energy within some finite volume. Thus, in order to de-excite dimensions one has to raise energy in some finite region of space. This can be achieved in heavy ion collisions where multiple particles collide, but not in two-particle collisions. An obvious problem is then that one has to raise the energy by four orders of magnitude in order to raise the temperature by one order of magnitude (since $E/V \sim T^4$), which would make the
$(3+1) \rightarrow (2+1)$ dimensional cross-over invisible at the LHC even if the crossover temperature is as low as $m_3 \sim 1$ TeV. This has  significant implications for the new physics at the LHC.  A clear prediction of this limit of temperature dependent number of dimensions is that the LHC will be practically blind to new physics even when working with its full power. Only slight deviations from the standard $(3+1)$-dim physics might be expected. A possible theoretical drawback of this extreme scenario is that the hierarchy problem might not be solved by temperature dependent physics, since the corrections to the Higgs mass will still be quadratically divergent in vacuum, up to the kinematic UV cut-off $N^{1/4}_c$. In contrast, the classical rigid network (or ordered lattice) limit would clearly solve the hierarchy problem, if $N^{1/4}_c \sim$ few TeV, since higher energy particles in vacuum propagate in lower dimensional space-time.

Of course, a generic situation is somewhere between these two extreme limits of energy vs. temperature dependent new physics. A smooth transition between these two limits may be observed in events with high multiplicity and also density (number of particles per unit volume).
The best place to look for experimental evidence are the cosmic rays experiments. Cosmic rays collide particles in our atmosphere with center of mass energies of $100$ TeV and more, which is high above the LHC energies.  More importantly, cosmic rays often produce very high multiplicity events with hundreds of particles in the single collision. Though it is very difficult to determine whether full thermal equilibrium had been established during the interaction, this regime is much closer to the high temperature environment than the events at the LHC. This might be the main reason why the earlier high altitude cosmic rays experiments observed planar propagation of secondary showers \cite{et1986,Mukhamedshin:2005nr,Antoni:2005ce}. It was noticed that only super-families with very high number of particles have planar alignment. The problem there was that very few super-families were observed, so it is still not clear if this effect was a statistical fluke or not. Current cosmic rays experiments are not performed at high altitude, so it seems very unlikely to replicate the results since energy of the shower degrades very quickly if one is not able to catch the very beginning of the shower at a high altitude.
The only exception here might be neutrinos.  Neutrinos interact weakly and unlike protons and photons can penetrate the whole depth of the atmosphere and interact for the first time in the detector so that the beginning of the shower can be caught.  Indeed, IceCube recently detected two PeV neutrino events which light up the whole detector by producing hundreds of particles \cite{Aartsen:2013bka}. Unfortunately, these events had the center of mass energy of only $1.4$ TeV, while the observed threshold for the planar events in earlier experiments was $4$ TeV.  It will be very important to collect events that originate in the detector and have above the threshold energy, and check the topology of the produced showers. If earlier observed alignment is also observed by IceCube, this might strongly support the model we discussed here.

One of the potential problems with this model of emerging dimensions might be possible Lorentz invariance violations in the light of strong Fermi constraints.   High energy photons propagating from a distant part of the universe toward us may be affected by discrete nature of space-time which in turn could modify the dispersion relation. Concretely, in the discrete ordered lattice limit, photons with the wavelength longer than the dimensional cross-over length scale would propagate in $(3+1)$-dim space-time, while those with the much shorter wavelength would see $(2+1)$-dime space-time. This may potentially lead to modified dispersion relation, or a time of arrival delay when two photons (one above and one below the cross-over scale) are compared. One of the ways to evade strong Lorentz invariance violations is to have a random lattice, or causal set, as discussed in \ref{el}, where Lorentz invariance violation would be stochastic and would average to zero, thus avoiding systematic violation of the dispersion relations. We also note that the two photons used by Fermi to put the constraints were both below TeV energies \cite{Vasileiou:2010nx}. They observed one $3$ GeV and one $31$ GeV photon coming to the detector with the time delay of about 1 second. However, if the crossover scale is set to a TeV by the solution to the hierarchy problem, none of these two photons actually probes the lower dimensional regime.
Obviously, individual high energy quanta propagating toward us from the other end of the universe do not propagate in the high temperature regime, and thus do not probe the thermal dimensional cross-over mechanism, discussed in Sec. \ref{sec:stringy}. For them to see a lower dimensional space-time, they would have to propagate through hot plasma with temperature higher than the cross-over scale. Thus, they always see $(3+1)$-dim space-time and Fermi constraints do not affect them at all.

Early universe observations like primordial gravitational waves would however be affected by lower dimensionality and could provide a very useful window into the dimensional history of our universe \cite{Mureika:2011bv}.  Most of the conventional lower-dimensional space-times have no propagating gravitational degrees of freedom, so gravitational waves cannot be produced in that epoch in the early universe. This can place a universal maximum frequency at which primordial gravitational  waves
can propagate. If the dimensional transition happened when the temperature of the universe was around a TeV, this cut-off frequency may be accessible to future gravitational wave detectors.

\section{Conclusions}

In conclusions, we formulated a model that captures the basic idea of ``evolving dimensions". Using the existing machinery of the string theory, where the coordinates in the target space $X^\mu$ can be viewed as fields defined on the worldsheet of a string $X^\mu (\xi^a)$, we added mass terms that break conformal symmetry of the worldsheet action in IR at finite temperature. In our construct, the masses decrease with the temperature of the environment. At the Big Bang, all of the ``dimensions'' are very heavy (of the order of the fundamental mass scale, perhaps $M_{Pl}$)  and are not excited. As the temperature drops, dimensions are excited one by one. This sequential opening of the dimensions allows for some fundamental problems in physics to be attacked from a completely new perspective.  We also presented a geometric/kinematic version of Brandenberger-Vafa argument for how  a discrete causal set, and eventually a $(3+1)$-dim Lorentz-invariant continuum space-time, along with Einstein gravity can emerge in the IR, from a $(1+1)$-dim worldsheet theory in the ultraviolet. $(3+1)$-dim space-time is special in this framework, as it has twice the dimension of the worldsheet, and can survive as a string network condensate, even after other dimensions open up.

The main problem with adding the mass terms for the fields $X^\mu (\xi^a)$ is the conformal invariance. If we want to interpret our model as an effective model of evolving dimensions, that conformal invariance is not an issue. However, in the context of the formal string theory, further justification is needed. We do note that in principle, a theory could be fundamentally conformal, but finite temperature effects can violate conformal invariance. In our case, at zero temperature, all the masses go to zero, and the theory is conformally invariant. Further, at any finite temperature, classical conformal invariance is restored in the subset of dimensions which are massless.  So, at any finite temperature an observer would quantize an ordinary string theory. Such an observer would notice that in order to cancel anomalies he needs more than (say) $(3+1)$ dimensions, so he would conclude that at lower energies additional dimensions must be excited (and presumably comes up with the model where the number of dimensions depend on temperature). This is analog with the current situation in the string theory where a $(3+1)$-dim observer concludes that at high energy additional dimensions should open up in order for anomalies to cancel. Unfortunately, this requires compactification with all of the problems that come with it. In the context of the formal string theory, ultimately one would want to obtain the definition of quantum observables in spacetime from the microscopic theory exactly, however, at this point we would like to leave this question for future work on the topic.

We then discussed phenomenological implications of our construct. Depending on the scales in the model, a dimensional cross-over can happen dynamically/thermally, at high enough temperatures, or kinematically, if particle energies exceed a characteristic threshold. For the kinematic dimensional cross-over, the standard model hierarchy problem  can be resolved by lowering the number of dimensions at the TeV scale, and also give distinct experimental signatures at the LHC.
In the opposite limit, if dimensions shut off at high temperatures (rather than high energies), then the LHC would be practically blind to any new physics. An intermediate picture between these two limits, would contain elements of both.

In contrast, cosmic ray experiments, especially those that can register the very beginning of the shower, and collisions with high multiplicity and density of particles, might be sensitive to {\it the thermal} dimensional cross-over.  Future gravitational wave detectors that can probe thermal history of our universe could also detect {\it the thermal} dimensional cross-over by registering a cut-off frequency above which there are no gravitational waves.

The model we presented here is unitary. The dimensions are not created out of nowhere, they always exist as degrees of freedom, but are not excited at high temperatures. The whole construct could be embedded in string theory,  with additional ingredients, within $10$ or $26$ dimensions. Then there would be no need for compactification, and large ambiguity in the choice of vacua,  or landscape could be avoided.

{\it Acknowledgement:} We are grateful to Ruth Gregory, Matt Johnson and John Hutchinson for helpful discussions. This work was supported by the Natural Science and Engineering Research Council of Canada, the University of Waterloo and Perimeter Institute for Theoretical Physics. Research
at the Perimeter Institute is supported by the Government of Canada through Industry
Canada and by the Province of Ontario through the Ministry of Research \& Innovation.
This work was also partially supported by the US National Science Foundation, under Grant No. PHY-1066278 and PHY-1417317.

\appendix
\section{Field Renormalization}

Here, we outline a geometric derivation of the Lorentz-invariant statistical averaging over string intersections, for parameter $Z$, defined as (Eq. \ref{z_def}):
\be
Z \equiv 8 \left\langle \int (\sqrt{-\gamma_1} d^2\xi_1)( \sqrt{-\gamma_2} d^2\xi_2) \frac{\delta^4(\Delta X^\mu)}{\sqrt{-g}}  \right\rangle.
\ee

As strings can be approximated as flat worldsheets near intersections, we can write:
\bea
X^\mu_1(\sigma,\tau) =  A_1^\mu \sigma +  B_1^\mu \tau, \\
X^\mu_2(\sigma,\tau) =  A_2^\mu \sigma +  B_2^\mu \tau.
\eea
Furthermore, we can go to the Minkowski coordinates, where the integral can be carried out to give:
\be\label{z_2}
Z= 8 \left\langle \frac{\sqrt{\left[A^2_1 B^2_1 - (A_1^\mu B_{1\mu})^2\right]\left[A^2_2 B^2_2 - (A_2^\mu B_{2\mu})^2\right]}}{|\epsilon_{\mu\nu\alpha\beta} A^\mu_1A^\nu_2B^\alpha_1B^\beta_2|}\right\rangle,
\ee
where $\epsilon_{\mu\nu\alpha\beta}$ is the Levi-Civita tensor in Minkowski space-time, while $A^2_i = \eta_{\mu\nu} A^\mu_i A^\nu_i$ and $B^2_i = \eta_{\mu\nu} B^\mu_i B^\nu_i$.

Using the Lorentz symmetries of the target space AND world sheet coordinates, without loss of generality, we can set:
\bea
A_1=(0,1,0,0),\\
B_1=(1,0,0,0) .
\eea
Furthermore, Eq. (\ref{z_2}) is invariant under re-scaling of each four-vector separately, so we can set $A^2$'s and $B^2$'s to -1 and +1 respectively (where we use $(+,-,-,-)$ signature).
Now, defining:
\be
A_2 = (a^0,{\bf a}), B_2 = (b^0,{\bf b}),
\ee
and using a Lorentz-invariant integration measure, we find:
\be
 Z= \frac{8 \int \frac{d^3a d^3 b}{\sqrt{(a^2-1)(b^2+1)}} \frac{\sqrt{1+\left(\sqrt{a^2-1}\sqrt{b^2+1}-{\bf a}\cdot {\bf b}\right)^2}}{|a_y b_z- a_z b_y|}}{\int \frac{d^3a d^3 b}{\sqrt{(a^2-1)(b^2+1)}}},
\ee
The integrals in numerator and denominator are dominated by $|{\bf a}|,|{\bf b}| \gg 1$. In this limit, they can be simplified to give:
\be
Z \simeq \frac{8 \int \frac{d^3a d^3 b}{|{\bf a}||{\bf b}|} \frac{|{\bf a}||{\bf b}|-{\bf a}\cdot {\bf b}}{|a_y b_z- a_z b_y|}}{\int \frac{d^3a d^3 b}{|{\bf a}||{\bf b}|}}.
\ee
Now, since the above expression is invariant under rescaling 3-vectors ${\bf a}$ and ${\bf b}$, we can reduce it to integrals over a unit sphere. After applying $SO(3)$ symmetries, the integrals reduce to:
\be
Z  \simeq \frac{1}{\pi} \int d\cos\theta d\cos\theta' d\phi \frac{1-\cos\theta \cos\theta'-\sin\theta \sin\theta' \cos\varphi}{|\sin\phi|  \sin\theta \sin\theta'},
\ee
where $0\leq \theta,\theta' \leq \pi$, while $0 \leq \varphi < 2\pi$. It is easy to see that only the first term is non-zero, and in fact logarithmically divergent as $\sin\varphi \rightarrow 0$:
\be
Z \simeq 4\pi \log\left(2\over \epsilon\right),
\ee
where
\be
\epsilon = 2 {\rm min} \left|\tan\left(\frac{\varphi}{2}\right)\right|.
\ee

The physical interpretation of this divergence is that the integral over the Dirac delta function blows up when the 2nd string moves nearly parallel to the 1st string, in its rest frame. However, given that the function $K(\Delta X^2)$ is resolved on the string length scale $\alpha'^{1/2}$, and the strings have a characteristic extrinsic curvature, ${\cal K}$, either due to thermal or quantum fluctuations, we expect a natural cut-off of:
\be
\epsilon \sim \varphi_{\rm min} \sim {\cal K}\alpha'^{1/2}.
\ee

\end{document}